\documentclass{optica-article}

\journal{opticajournal} 
\articletype{Research Article}

\newcount\Comments  
\usepackage{changes}
\newcommand{\note}[2]{\ifnum\Comments=1\textcolor{#1}{#2}\fi}
\definechangesauthor[name={Author 1}, color=red]{author1}

\usepackage{lineno}

\begin{document}

\title{Intrinsic Temporal and Spectral Mixing
 in Time-Resolved Terahertz Spectroscopy}

\author{Benjamin J. Dringoli and David G. Cooke\authormark{*}}

\address{Department of Physics, McGill University, Montreal, QC, Canada H3A2T8}

\email{\authormark{*}dave.cooke@mcgill.ca}

\begin{abstract*} 
In an ultrafast optical-pump terahertz-probe measurement, the photoinduced material response can be modulated on a timescale shorter than the extent of the THz pulse. In this situation, the measured time-frequency response deviates from a simple time-dependent linear response. When full two-dimensional time-frequency maps are measured, this yields complex features that can be incorrectly assigned to a photoexcited coherent response. We investigate this experimentally via the measured response of photoexcited SnSe, whereby photoinduced phase change dynamics lead to ultrafast changes of the charge carrier and lattice optical conductivity response. Two-dimensional time-frequency THz transmission maps subsequently show unexpected time-frequency features at early pump-probe delay times. These features are reproduced in both finite-difference time-domain simulations of the THz experiment and in an extension of non-equilibrium response function theory, demonstrating their systematic origin. This work improves the understanding of systematic effects in high time resolution optical-pump THz-probe spectroscopy, and explores the conditions in which they are likely to appear.
\end{abstract*}

\section{Introduction}

\indent Time-resolved terahertz (THz) spectroscopy (TRTS) has proven itself a powerful method for measuring ultrafast conductivity dynamics with sub-picosecond temporal resolution, beyond the capability of conventional electronic methods \cite{JepsenLPR2011}. In this technique, phase stable single-cycle pulses of THz frequency light are used as probe pulses to interrogate a material system following the absorption of a femtosecond-duration optical pulse. Full response function retrieval requires time-sampling of the affected THz transients, typically through free-space sampling in an electro-optic crystal, for various pump-THz probe delay times \cite{JepsenLPR2011,KindtJPC1996}. Subsequent Fourier transformation along the probe time axis yields complex-valued amplitude and phase spectra at each pump-probe delay, which can be transformed into time-dependent linear response functions such as the complex conductivity $\sigma(\omega,\tau)$ or dielectric function $\epsilon(\omega,\tau)$  \cite{UlbrichtRMP2011}.

\indent The methodology of TRTS has been extensively studied, focusing on optimal pulse delay configurations \cite{KindtJCP1999}, analytic current models \cite{NemecJCP2002,Nemec2JCP2005}, and detector response function effects \cite{LeitenstorferAPS1999,JepsenLPR2011,LarsenJOSAB2011}. As THz spectroscopy measures the current response to the applied THz field imprinted onto that THz field to detect material properties, the extraction of optical response functions relies upon current response linearity assumptions. For photoexcited materials that exhibit strong changes on timescales less than the duration of the THz pulse, this linear link between the applied THz field and the subsequent current response can be broken when the pump and probe overlap in time and space. The effect of these modified current responses on spectroscopy results were initially investigated in the context of superconducting materials \cite{OrensteinPRB2015}, where long momentum relaxation times cause the distorted current response to influence spectroscopy results even at times after pump-probe overlap. This affected photocurrent induces time-frequency artifacts in the reconstructed material response functions if not properly accounted for, and was explored in the case of a Drude-like photoexcited conductivity spectrum.

\indent As was realized by Dodge and Orenstein in Ref.~\cite{OrensteinPRB2015}, any material that is strongly perturbed on short timescales and exhibits a long momentum relaxation time should exhibit similar artifacts in recorded THz spectra. In the following work, we extend the investigation of early-time non-equilibrium responses in THz measurements to a wider range of photoexcited materials using finite-difference time-domain (FDTD) simulations \cite{LarsenJOSAB2011} and an extended case of the non-equilibrium response model proposed in Ref. \cite{OrensteinPRB2015}. We confirm that these effects arise only when the rise time of the photoconductivity is short and there exists a long momentum scattering time, and can produce spectroscopy features that may be misinterpreted as coherent processes. The simulated results are compared to experimental data showing unexpected time-frequency features in the recorded THz optical conductivity spectra of SnSe, a strongly polar material which exhibits ultrafast changes to both phonon and free-carrier responses upon photoexcitation. While the detailed physics of photoexcited SnSe is discussed in another publication, here we focus on the early-time systematic effects present in such THz experiments. This is particularly relevant as THz spectroscopy is used to study a wider range of material responses such as photoinduced phase transitions \cite{DringoliPRL2024,XueJAP2013} or fast-acting coherent polarization transients \cite{ZhangAS2021}.

\begin{figure*}[t!]
    \centering
    \includegraphics[width=\textwidth, trim={1.5cm 6cm 1.5cm 5cm},clip]{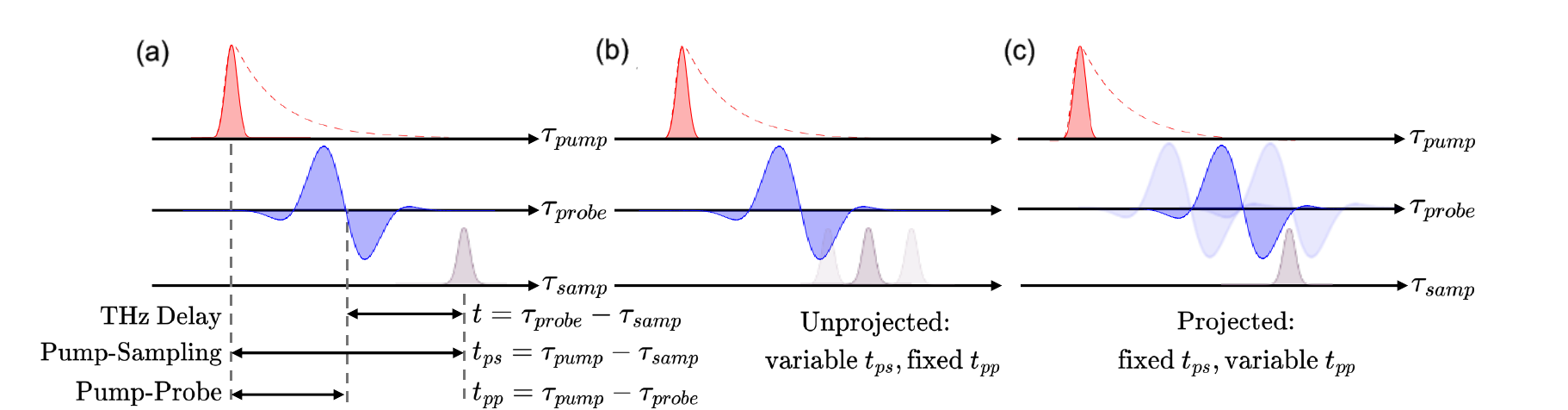}
    \caption{(a) Pulse diagram showing the absolute ($\tau_{pump}$, $\tau_{probe}$, $\tau_{samp}$) and relative ($t$, $t_{ps}$, $t_{pp}$) time axes for a TRTS experiment involving pump (top), probe (middle), and sampling (bottom) pulses. (b) Pulse delay representation showing the `unprojected' delay scheme where $t_{ps}$ is varied to collect THz waveform data. (c) Similar representation showing the `projected' case where $t_{ps}$ is fixed and $t_{pp}$ is varied. The SnSe experiments discussed in this work use the delay scheme shown in (c), where each part of the sampled THz probe experiences the same pump delay, whereas the simulation results model the scheme shown in (b).}
    \label{Artfig1}
\end{figure*}

\section{Materials and Methods}

\begin{figure}[t!]
    \centering
    \includegraphics[width=0.95\textwidth, trim={0cm 0cm 0cm 0cm},clip]{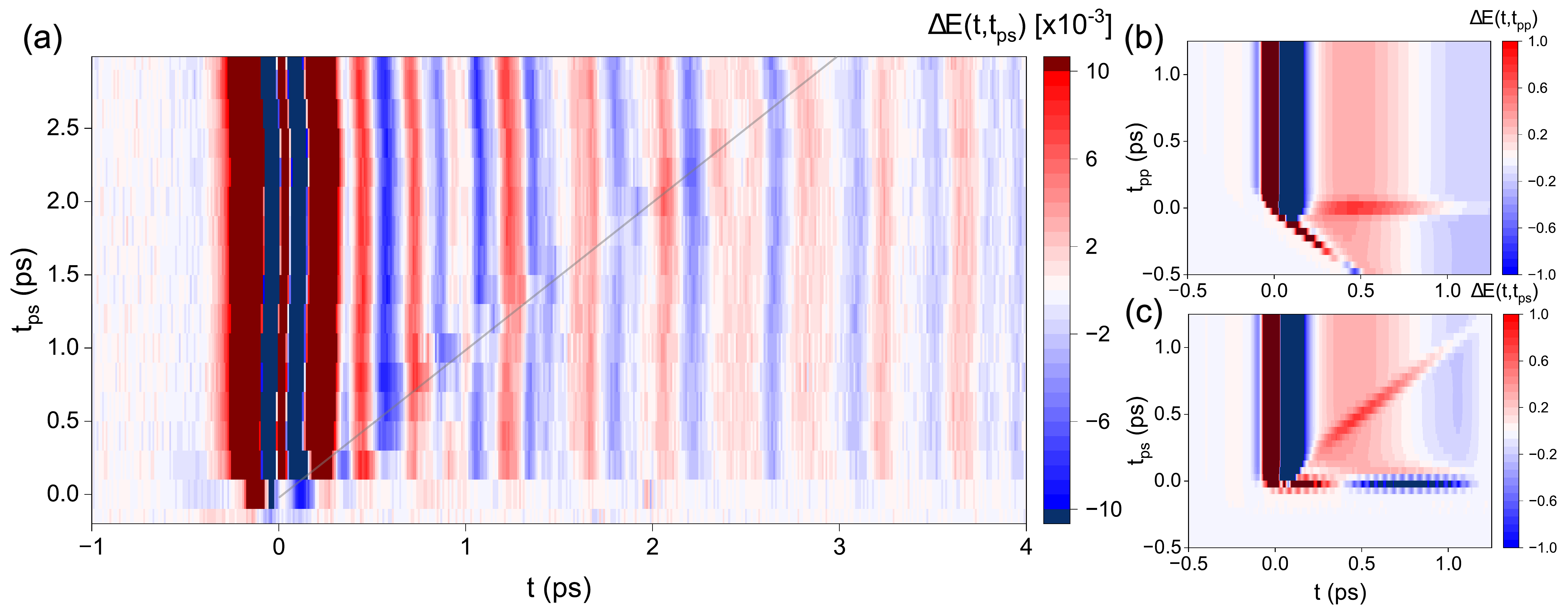}
    \caption{(a) Experimental time-domain photoinduced change in THz field $\Delta E (t,t_{ps}) = E_p(t,t_{ps}) - E_r(t)$ with guide line showing an additional feature along $t = t_{ps}$. (b) Unprojected FDTD simulation of a THz experiment showing a long-lived response at $t_{pp} = 0$, where the pump arrival crosses the main peak of the THz pulse. (c) Projection of the data in (b) from $(t,t_{pp})$ to $(t,t_{ps})$ showing how the projected $t_{pp} = 0$ feature now follows the $t = t_{ps}$ line, matching that seen for the measured data in (a). All color scales are reduced to 10\% of the full signal range.}
    \label{Artfig2}
\end{figure}

\indent The ultrafast THz photoconductivity spectrum of SnSe was measured via TRTS using air-plasma THz generation \cite{KimOE2007} and air-biased coherent detection \cite{DaiPRL2006,DAngeloOSA2016,HoOE2012}. Transmitted single-cycle multi-terahertz pulses with frequency bandwidths of 0.5-12~THz were sampled in time after the arrival of a 35~fs, 800~nm optical pump pulse. The different laser pulses and related time axes used for TRTS are shown in Figure \ref{Artfig1}(a). The choice of delay staging used for these TRTS experiments is relevant for the following discussion: TRTS can be recorded with a fixed pump-probe delay $t_{pp}$ and variable sampling delay $t_{ps}$, as shown in Figure \ref{Artfig1}(b), but this leads to different parts of the THz probe pulse experiencing different material responses as the material evolves after excitation \cite{Nemec2JCP2005}. The THz spectrometer used here instead employs the delay staging from Kindt et al. \cite{KindtJCP1999}, shown in Figure \ref{Artfig1}(c), where the pump-sampling delay $t_{ps}$ is fixed. This allows for the retrieval of two-dimensional time-time maps (Figure \ref{Artfig2}(a)) for which all components of each reconstructed THz probe pulse have experienced the same pump-probe delay $t_{pp}$ across the entire probe time $\tau_{probe}$.  

\indent The measured SnSe sample was a $2.5$~mm $\times$ $2.5$~mm $\times$ ${\sim}500$~nm exfoliated single crystal grown using methods described in Ref.~\cite{ZhaoNAT2014}, and from the same parent crystal as in Refs.~\cite{ZhaoNAT2014,ReneDeCotretPNAS2022,DringoliPRL2024}. The exfoliated crystal was mounted on a 500 $\mu$m-thick electronic-grade diamond substrate, used for its constant THz refractive index, low optical absorption, high thermal conductivity, and negligible photoconductivity from shallow boron impurities (here ppb). We have verified that the substrate alone does not show any photoconductive response under the same excitation conditions. The SnSe-on-diamond was placed in a sample-in-vacuum cryostat (Janis ST-300MS) and measured via transmission at normal incidence with the THz and pump pulses linearly polarized along the SnSe c-axis. The TRTS data discussed here was recorded at a sample temperature of 80~K and with pump fluences between 0.1 and 7.5 mJ/cm$^2$.

\indent The transmitted THz electric field was sampled in time $t$ using a double-modulation scheme to extract the THz transmission with optical excitation $E_p(t,t_{ps})$ and the optical pump-induced differential between pumped and reference fields $\Delta E(t,t_{ps}) = E_p(t,t_{ps}) - E_r(t)$ at various pump delay times $t_{ps}$. The complex pump-induced transmission function given as
\begin{equation}
\begin{split}
    T(\omega,t_{ps}) = \frac{|E_p(\omega,t_{ps})|}{|E_r(\omega)|}
    \big[ \textup{cos}(\phi(\omega,t_{ps}))+ i \ \textup{sin}(\phi(\omega,t_{ps})) \big],
\end{split}
\end{equation}
where the respective amplitudes and phase change $\phi = \phi_p - \phi_r$, can be extracted via Fourier analysis and transformed into other response functions such as $\sigma(\omega,t_{ps})$.

\indent An FDTD formalism simulating a TRTS experiment was used to explore the early time response, with details outlined in Ref.~\cite{LarsenJOSAB2011}. Sample dispersion is handled via the auxiliary differential equation method for Lorentzian oscillators (phonons), with Drude free carriers introduced by optical photoexcitation at a wavelength of 800~nm (1.55~eV), above SnSe's indirect band gap energy of 0.9 eV. A pump penetration depth of 60~nm was chosen to match that of SnSe at 800~nm \cite{MakinistianPSSB2009}. FDTD allows for both excitation and material parameters (pump time-width $\Delta \tau_p$, carrier density $n$, phonon center frequency $\omega_{TO}$, phonon lifetime $\Gamma_{TO}$) to be independently controlled, allowing for relevant parameter scaling to be explored. Here subsequent diffusion of charge carriers is ignored as we focus on the earliest times following photoexcitation. To model the experiment, transmitted THz waveforms are propagated in the presence and absence of a pump pulse, and the calculated differential THz waveform $\Delta E(t,t_{pp})$ is collected into a two dimensional time-time map (Figure \ref{Artfig2}(b)). This response can then be numerically projected into $\Delta E(t,t_{ps})$ to match the measured THz response from the delay staging used in the TRTS experiments (Figure \ref{Artfig2}(c)) \cite{JepsenLPR2011}.

\indent It is important to note that the recovered fields in FDTD are true fields without any influence of a detector response function, typically present when detecting THz via electro-optic sampling inside a detection crystal. Not including such a response function will approximate a situation where the entire bandwidth of the THz source is within a non-dispersive region of the THz detector. For air-biased coherent detection with $\sim$35~fs sampling pulses, the detector response function is approximately flat for frequencies up to and beyond 10~THz \cite{WangJIMTW2016}. As propagation effects through the sample only influence the absolute phase of the detected pulses and we are interested in differential modulations, we directly compare the fields from the simulations to the experimental data.

\section{Results and Discussion}

\indent Figures \ref{Artfig2}(b,c) show time-domain FDTD results of a TRTS experiment performed in transmission near $t=t_{pe}$ and $t = t_{ps}$, respectively. In the unprojected $\Delta E(t,t_{pp})$ map in Fig. \ref{Artfig2}(b), the pump arrival time and its influence on the THz pulse varies linearly across $t$, seen in the $t_{pp} = -0.5$ to 0~ps region. This pump-induced change map clearly demonstrates how in the unprojected case, different parts of the THz probe can experience different effective pump delays near the pump arrival time. Fig. \ref{Artfig2}(c) shows the same results after a numerical projection of the data in Fig. \ref{Artfig2}(b) along $t = t_{pp}$. This aligns the distributed pump arrival times onto $t_{ps} = 0$, matching the method used in experiments to record the SnSe data with consistent pump delays across the entire THz pulse.

\indent The focus of the discussion here is the presence of an extended feature at the pump-probe overlap time ($t_{pp} = 0$) in the pre-projected representation. This feature, upon projection from $t_{pp}$ to $t_{ps}$, is shifted upwards into the post-excitation region of the data along $t = t_{ps}$. By reducing the range of the color scale in the experimental $\Delta E(t,t_{ps})$ data shown in Figure \ref{Artfig2}(a), a clear response enhancement following $t = t_{ps}$ becomes visible emanating from the overlap point between the THz probe and the 800~nm, 35~fs optical pump, matching the behavior seen in the FDTD simulation results.

\indent The origin of this feature can be explained by considering the material current response to the applied THz field and how it changes with optical excitation. For a generalized time $t$, the recovered THz probe effectively measures the material's current response to the THz field $E(t)$ after photoexcitation $I(t)$ as

\begin{equation}
    \delta J (t) = \iint_{-\infty}^{\infty} E(t - t') I(t - t'') \sigma^{(3)}(t',t'') dt' dt''.
\end{equation}

\begin{figure}[t!]
    \centering
    \includegraphics[width=\textwidth, trim={0cm 0cm 0cm 0cm},clip]{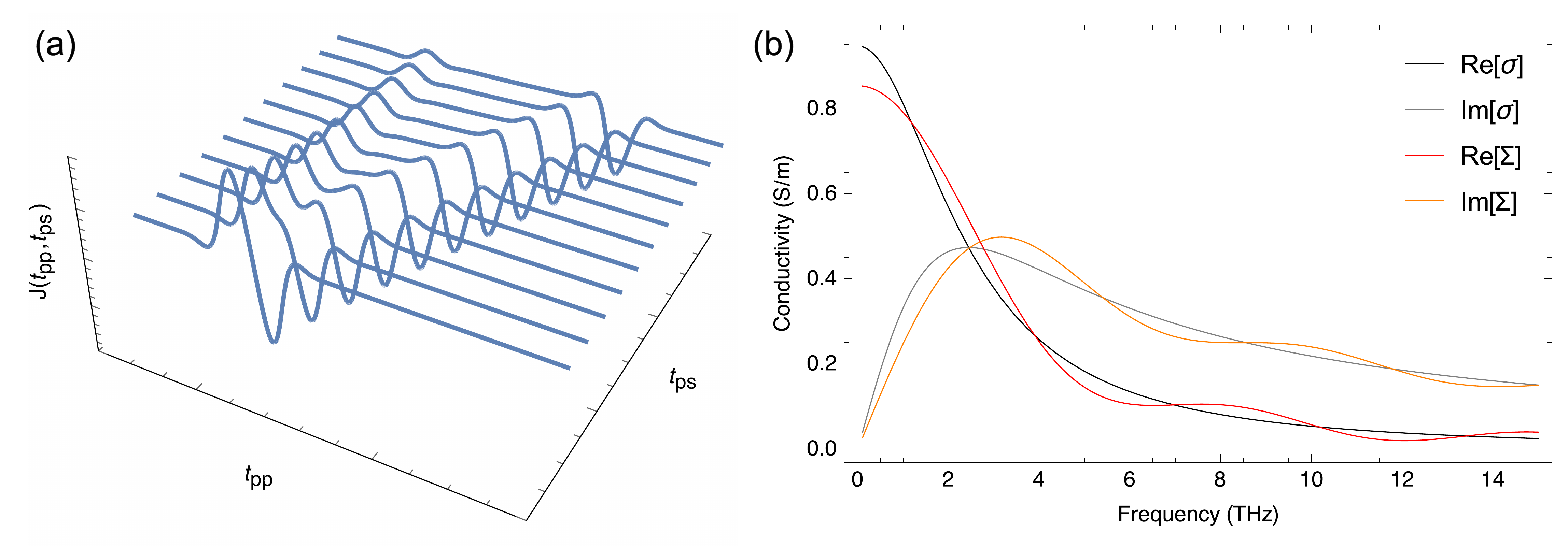}
    \caption{(a) Nonlinear current response as given by Eqn. \ref{eqn1} for various pump-sampling delays $t_{ps}$ and $\gamma$ = 0.3 ps$^{-1}$. Note the extension of the current response to later times as the pump-sampling delay is increased. (b) Normalized comparison between typical Drude response in the THz regime $\sigma$ and the nonlinear current model given by Equation \ref{eqn2} $\Sigma$. At early sampling times ($t_{ps} \approx 1$~ps) the extended nonlinear current response leads to additional time-frequency oscillations that are visible in spectroscopy results. Here $\gamma$ has been set equal to 2 to better display the modulations.}
    \label{Artfig3}
\end{figure}

This formulation of the measured current response can be simplified by defining a two-dimensional third order response function $\Sigma(t,t') = \int_{-\infty}^{\infty} I(t - t'') \sigma^{(3)}(t',t'') dt''$ \cite{OrensteinPRB2015}, leaving the change in current response equation in typical linear-response form

\begin{equation}
    \delta J (t) = \int_{-\infty}^{\infty} E (t-t') \Sigma(t,t') dt',
\end{equation}

\noindent but with a response function implicitly tied to $I(t)$ and lacking time invariance. This $\Sigma$ form is what is effectively measured in TRTS \cite{KindtJCP1999}, and in circumstances close to the pump-probe overlap time, may deviate from the true optical conductivity response $\sigma$.

\indent When the THz pulse overlaps with the pump excitation in time, only a latter portion of the THz pulse experiences the photoexcited medium, as shown in the $t_{pp} < 0$ region of Figure \ref{Artfig2}(b). This change in the material's response within the time extent of the THz probe pulse leads to a current response that is not linear to the input THz field. This effect can be seen in the expression of the nonlinear current response in TRTS experiments considering a Drude impulse response function (Eqn. 12 in \cite{OrensteinPRB2015}):

\begin{multline}
    \delta J (t_{ps},t_{pp}) = \Theta(t_{ps}) \frac{\delta n (0) e^2}{m} \textup{exp}(-\Gamma t_{ps}) \times \\ \int^{\infty}_{-\infty} E_{THz}(t_{ps} - t_{pp} - \tau) \times \Theta(t_{ps} -\tau) \Theta(\tau) \textup{exp}(- \gamma \tau) d\tau.
    \label{eqn1}
\end{multline}

The term before the integral represents the pump excitation and subsequent relaxation set by the recombination rate $\Gamma$. The integrand represents the current interaction, dependent on the difference between the probe and sampling delays (so either could be varied as discussed above) as well as the Drude scattering rate $\gamma$. The results of this expression for different values of $t_{ps}$ are shown in Figure \ref{Artfig3}(a). With the inclusion of a time-invariance breaking pump excitation term of the form $\Theta(t_{ps}) \frac{\delta n (0) e^2}{m} \textup{exp}(-\Gamma t_{ps})$, the current expression becomes explicitly dependent on $t_{ps}$ and $t_{pp}$ independently, instead of only their difference as with equilibrium THz time-domain spectroscopy (TDS). Taking into account the specific delay staging used to equalize the pump arrival time by fixing the pump-sampling delay $t_{ps}$, the extracted current response spectrum $\Sigma(\omega,t_{ps})$ can be solved for (Eq. 13 in \cite{OrensteinPRB2015}):

\begin{equation}
    \Sigma(\omega,t_{ps}) = \Theta(t_{ps}) \frac{\delta n (0) e^2}{m} \textup{exp}(-\Gamma t_{ps}) \frac{1}{\gamma - i \omega} \\ \times [1 - \textup{exp}(-\gamma t_{ps}) \textup{exp}(i \omega t_{ps})].
    \label{eqn2}
\end{equation}

\noindent The terms before the square brackets represent the typical instantaneous Drude conductivity spectrum in a TRTS experiment. The additional term in square brackets is what leads to additional time-frequency structure in recorded spectra after Fourier transforming the time-time data into a spectroscopic representation \cite{KindtJCP1999}. An example of the effect of this additional term compared to a typical Drude spectrum is shown in Figure \ref{Artfig3}(b).

\begin{figure}[t!]
    \centering
    \includegraphics[width=0.8\textwidth, trim={0cm 0cm 0cm 0cm},clip]{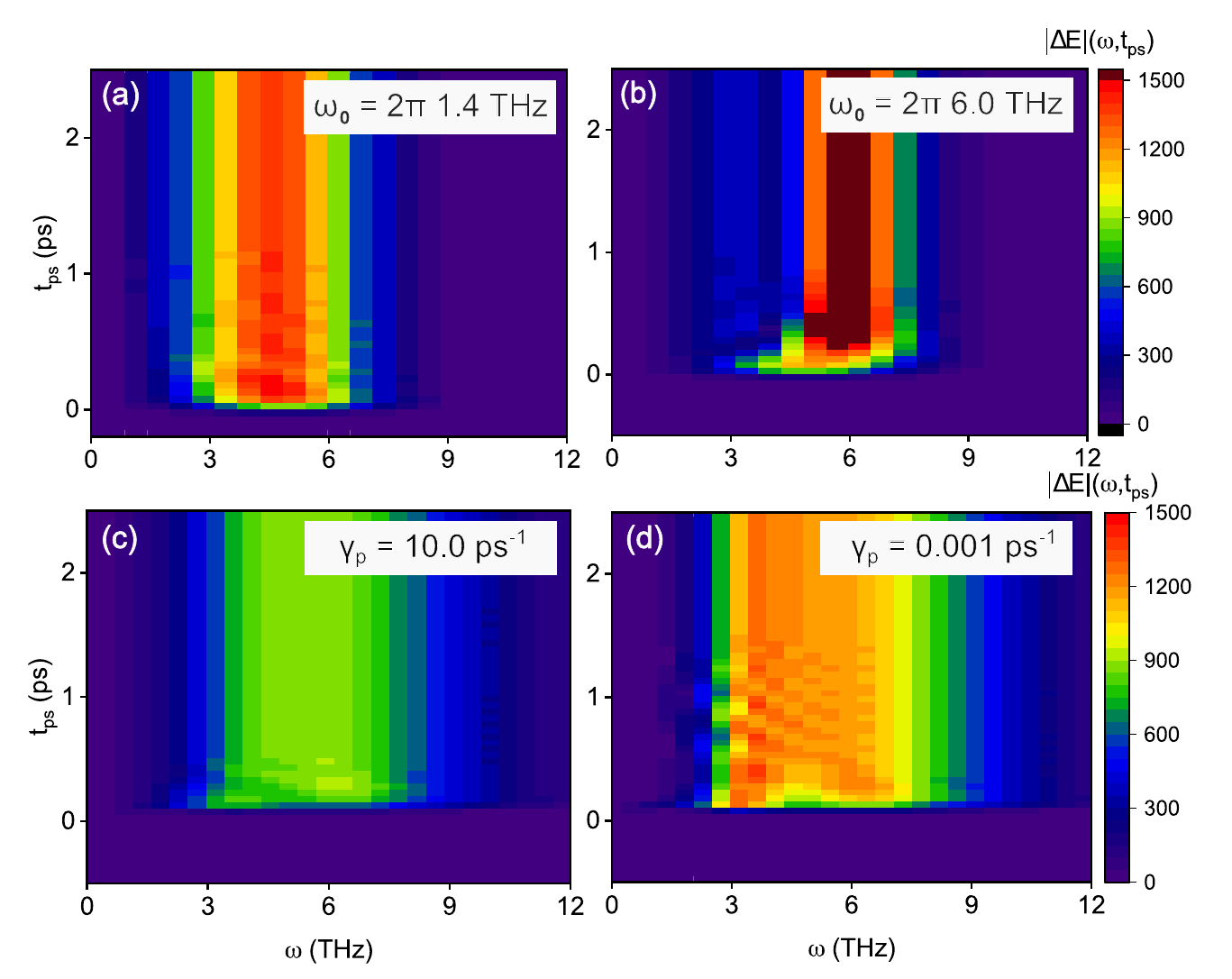}
    \caption{FDTD simulations of a THz transmission experiment on a material with phonon dispersion, showing time-frequency structure appearing in $\Delta E (\omega,t_{ps})$. Spectral oscillations appear centered around the simulated phonon frequency, shown for values of $\omega_0 =$ 1.4~THz (a) and 6.0~THz (b). Modifying the momentum scattering time of the photoexcited carriers shows that these features are less severe for short momentum scattering times (c) and are enhanced as the momentum relaxation time becomes long (d), as expected from the form of Equation 5.}
    \label{Artfig4}
\end{figure}

\indent To test how these effects could manifest in the case of TRTS on SnSe, Figures \ref{Artfig4}(a-d) show the recovered $\Delta E(\omega,t_{ps})$ magnitude in FDTD simulations of a THz transmission experiment with different material parameters. The FDTD simulations took into account the background Lorentzian dispersion of an optical phonon with Drude carrier photoexcitation provided by an optical pump pulse. Figs. \ref{Artfig4}(a,b) show that the time-frequency features move to be centered around the frequency region with largest dispersion, in this case the optical phonon response. Figs. \ref{Artfig4}(c,d) show that the oscillations are less severe for short momentum scattering times (0.1 ps) in comparison to an effectively infinite scattering time (1000 ps), as expected from the exp($-\gamma t_{ps}$) term in Equation 5.5. The time-frequency features are additionally suppressed in simulations employing longer duration pump pulses ($\Delta \tau_p \geq$ 100~fs), as shown in Supplemental Figure 1. This aligns with the non-equilibrium current interpretation, as with longer-duration pump excitations the change to the environment experienced by different parts of the THz pulse near the pump-probe overlap time is less abrupt, and therefore contains less high frequency Fourier components. This was also noted in the original non-equilibrium response function work \cite{OrensteinPRB2015}, where artifact oscillations are noted to be suppressed for $\gamma \times t_{ps} > 1$ and $\omega \times \Delta \tau_p > 1$.

\begin{figure}[t!]
    \centering
    \includegraphics[width=0.95\textwidth, trim={0cm 0cm 0cm 0cm},clip]{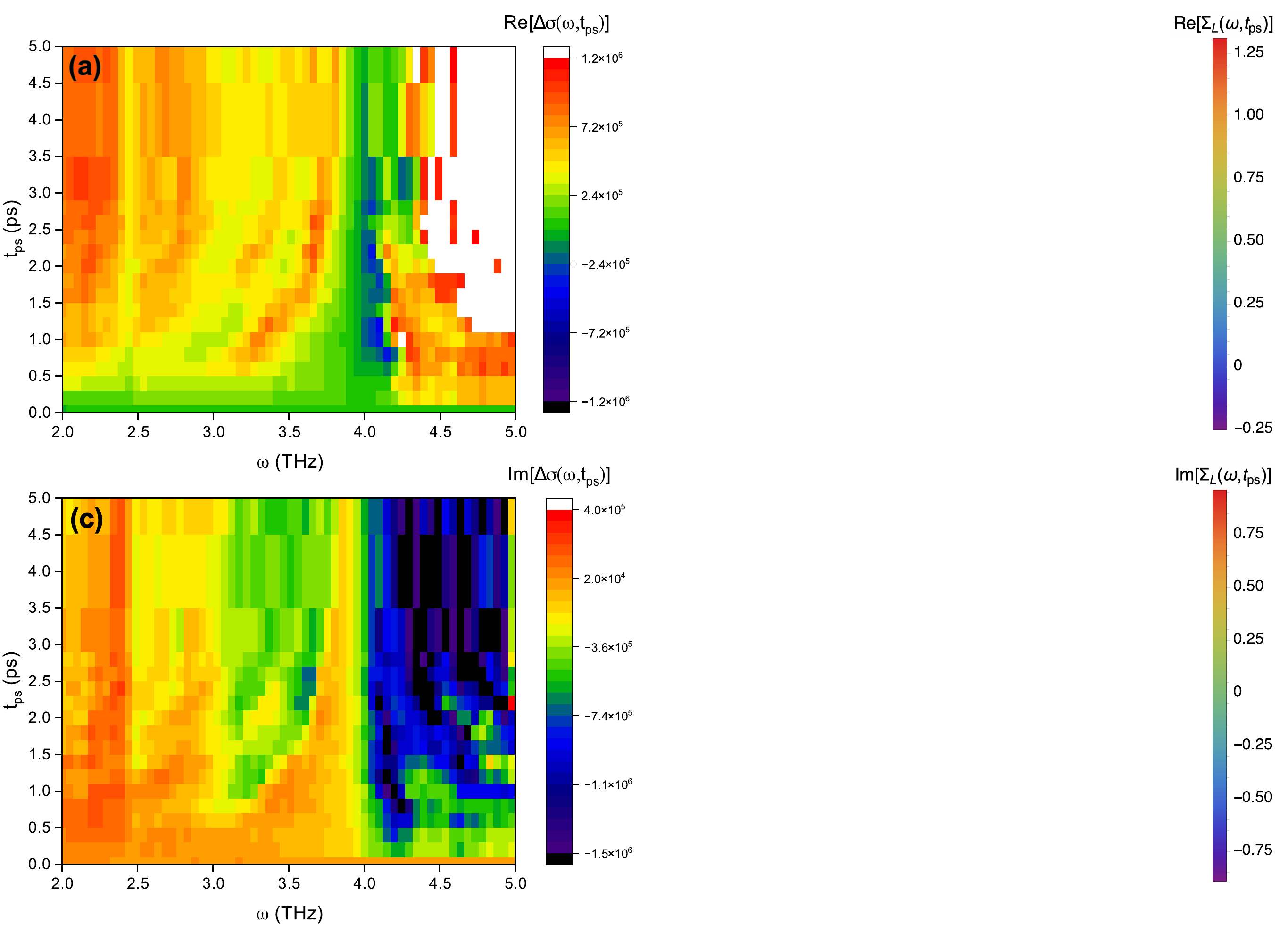}
    \caption{Comparison between the measured SnSe (a) Re[$\Delta \sigma(\omega, t_{ps})$] and (c) Im[$\Delta \sigma(\omega, t_{ps})$] versus the real (b) and imaginary (d) components of the derived nonlinear Lorentzian response $\Sigma_L(\omega;t_{ps})$. $\Sigma_L(\omega;t_{ps})$ reproduces both the dynamic time-frequency features and the Im[$\Delta \sigma(\omega, t_{ps})$] sign change seen in the SnSe TRTS data. The $\Sigma_L(\omega;t_{ps})$ shown here was calculated using $\omega_0$ = 4 THz, $\Gamma$ = 0.1 ps$^{-1}$, and $\gamma$ = 0.1 ps$^{-1}$.}
    \label{Artfig5}
\end{figure}

\indent In the case of SnSe, the main TO phonon near 3.9~THz has a linewidth of approximately 0.21~THz \cite{EfthimiopoulosPCCP2019}, leading to a current relaxation time of approximately 5~ps. This is well within the range where early-time oscillatory responses are expected in simulations and indeed observed experimentally. Figures \ref{Artfig5}(a,c) show the effects of the $\Delta E (t,t_{ps})$ enhancement in Figure \ref{Artfig2}(a) on spectroscopy following Fourier transformation along the $t$ axis. The experimental $\Delta \sigma(\omega, t_{ps})$ maps exhibit clear time-frequency features on either side of the main SnSe TO phonon response at 3.9~THz \cite{EfthimiopoulosPCCP2019} (at a sample temperature of 80~K), like those recovered from the FDTD simulations. The extent of these features with increasing $t_{ps}$ coincide with the presence of the artifact in the recorded time-time data shown in Fig. \ref{Artfig2}(a), and the feature scales positively with excitation fluence, shown in Supplemental Figures 2 and 3.

\indent The recovery of early-time time- and frequency-domain features in FDTD simulations demonstrates that these artifacts are intrinsic to any TRTS experiment when measuring materials soon after strong ultrafast excitation. Though, the form of the simulated features does not match well with those measured in SnSe experiments. This is attributed to a markedly non-Drude photoexcited response in SnSe, where photoinduced phase dynamics prompt an ultrafast modification to the recorded TO phonon spectrum. The previously-described non-equilibrium response formalism can therefore be extended by using the Lorentzian impulse response function:

\begin{equation}
   \textup{IRF}_\textup{L} =  \Theta(t_{pp}) \textup{exp} \left(-\frac{\Gamma t_{pp}}{2} \right) \left( \textup{cos}(\Omega t_{pp}) - \frac{\gamma}{2 \Omega} \textup{sin}(\Omega t_{pp}) \right),
\end{equation}

\noindent where $\Omega = (\omega_0^2 - \gamma^2/4)^{1/2}$ is the adjusted oscillation frequency. This impulse response can then be, like in the Drude case, substituted into the non-equilibrium expression to derive $\Sigma(\omega;t_{ps})$ for Lorentzian dispersion. This leads to a Lorentzian non-equilibrium response function $\Sigma_L$ which also contains an additional oscillatory term (similarly shown in square brackets):

\begin{multline}
    \Sigma_L(\omega;t_{ps}) = \Theta(t_{ps})\frac{\delta n (0) e^2}{m} \textup{exp}\left(- \Gamma t_{ps}\right) \left( \frac{\omega}{i(\omega_0^2-\omega^2)+\omega \gamma} \right) \times \\ \left[ 1+ \textup{exp}\left(\frac{-\gamma t_{ps}}{2}\right) \textup{exp}(i \omega t_{ps}) \left( \left(i \frac{\omega_0^2}{\omega \Omega}+\frac{\gamma}{2 \Omega} \right) \textup{sin}(\Omega t_{ps}) - \textup{cos}(\Omega t_{ps}) \right) \right].
    \label{Wakefield}
\end{multline}

\indent $\Sigma_L$ maintains many additional similarities to the Drude case, with $\Theta(t_{ps})$, $\textup{exp}(- \Gamma t_{ps})$, and the original Lorentzian dispersion all appearing before the additional time-frequency oscillation term. With these similarities it is expected that oscillations will also occur in addition to the input Lorentzian response in similar pumping and probing conditions. Figures \ref{Artfig5}(b,d) show the real and imaginary components of $\Sigma_L(\omega;t_{ps})$ evaluated with parameters accurate to the phonon and charge relaxation properties in SnSe ($\omega_0$ = 4 THz, $\Gamma$ = 0.1 ps$^{-1}$, and $\gamma$ = 0.1 ps$^{-1}$). The additional oscillations caused by the bracketed nonequilibrium term in Eq.~\ref{Wakefield} show much better qualitative agreement to those measured in the SnSe TRTS experiments, reproducing the `wakefield' dynamics with increasing $t_{ps}$ and the sign change in the imaginary component. This response moves with $\omega_0$ and is suppressed for large values of $\gamma$ and $\Gamma$, as shown in Supplemental Figure~4. The asymmetric central peak in the measured Re[$\Delta \sigma(\omega, t_{ps})$] map is likely caused by influence from the large optical conductivity enhancement above the main TO phonon response \cite{DringoliPRL2024}. This phonon response appears to demonstrate lower $\Delta \sigma$ than its surroundings due the influence of the background dispersion in the photoexcited measurement, but the presence of the time-frequency structure and its similarity to $\Sigma_L(\omega;t_{ps})$ is still clear.

\indent The excellent qualitative agreement between $\Sigma_L(\omega,t_{ps})$ and the measured SnSe TRTS response, as well as their coincidence with an extended response in the time-time maps, supports the interpretation that such early-time features are not real responses from coherent phonons or other complex pump-induced processes, but merely artifacts from probing fast, strong, narrow-linewidth excitations with a high bandwidth and resolution THz instrument. This dependence on long momentum relaxation time is additionally supported by suppressed time-frequency features in room temperature measurements with broader phonon responses, shown in Supplemental Figure 5. The presence of oscillatory features in all FDTD time-time maps regardless of the photoexcited response (Drude or Lorentz) highlights the intrinsic nature of these effects when employing pump pulses with duration much shorter than the THz probe duration. Despite this ubiquity, we are unaware of other experimental material studies which have noted the presence of these effects, which may be related to differences in the THz detection process. As again noted in the work by Dodge and Orenstein, without the benefit of largely dispersionless air-biased coherent detection, precise deconvolution processes are required to accurately reconstruct $\delta J (t_{ps},t_{pp})$ from the sampled THz field \cite{OrensteinPRB2015}. We believe this important intrinsic time-frequency artifact has been neglected to date, perhaps simply because of the particular experimental requirements and detailed analysis required to recognize it.

\section{Conclusions}

\indent In conclusion, high-resolution time-time and time-frequency maps provided by air-plasma-based TRTS have enabled investigations into the accuracy of extracted material response functions at early pump-probe delay times. In the case of SnSe, strong changes in the optical conductivity of narrow-linewidth phonon dispersion features allow for non-equilibrium distortions to be directly observed in experiments. FDTD simulations show the presence of oscillatory features in 2D THz time-frequency conductivity maps, behaving analogously to those predicted for nonlocal currents in high mobility semiconductors and superconductors \cite{OrensteinPRB2015}.

\indent Analytical expressions for the expected measured response function in the case of an excited Lorentzian response $\Sigma_L(\omega,t_{ps})$ demonstrate excellent qualitative agreement with the measured SnSe $\Delta \sigma(\omega, t_{ps})$ maps. These results show the importance of considering the possible systematic origins of exotic features in THz spectroscopy, especially as this technique is used to probe effects such as photoinduced phase transitions which involve strongly modulating materials' charge and phonon responses on ultrafast timescales. This analysis can likely be further extended to cover additional material responses or combinations thereof to better understand the intrinsic limitations of accurate response function extraction from TRTS experiments. \\

\begin{backmatter}

\bmsection{Acknowledgment} The authors thank the NSERC Discovery and Mitacs programs for funding this work.

\bmsection{Disclosures} The authors declare no conflicts of interest.

\bmsection{Data availability} Data underlying the results presented in this paper are not publicly available at this time but may be obtained from the authors upon reasonable request.

\bmsection{Supplemental document} See Supplement for supporting content. \\

\end{backmatter}

\bibliography{THzArtifactRefsFinal}

\end{document}


\title{Supplemental Information: Intrinsic Temporal and Spectral Mixing in Time-Resolved Terahertz Spectroscopy}

\author{Benjamin J. Dringoli,\authormark{1} David G. Cooke,\authormark{1,*}}

\address{\authormark{1}Department of Physics, McGill University, Montreal, QC, Canada H3A2T8}

\vspace{1cm}

\noindent Additional FDTD simulations were performed to test the dependence of various parameters. Supplemental Figure 1 shows the scaling of the time-frequency features with increasing pump pulse duration. The features are present for all simulations with pump pulse durations shorter than 100~fs. This links the observance of features with the speed of the photoinduced change in materials, which aligns with the non-equilibrium theory as shorter pump pulses create a strong change with extent less than the duration of the THz probe pulse.

\begin{figure}[ht!]
    \centering
    \includegraphics[width=0.9\textwidth, trim={10cm 0cm 11.5cm 0cm},clip]{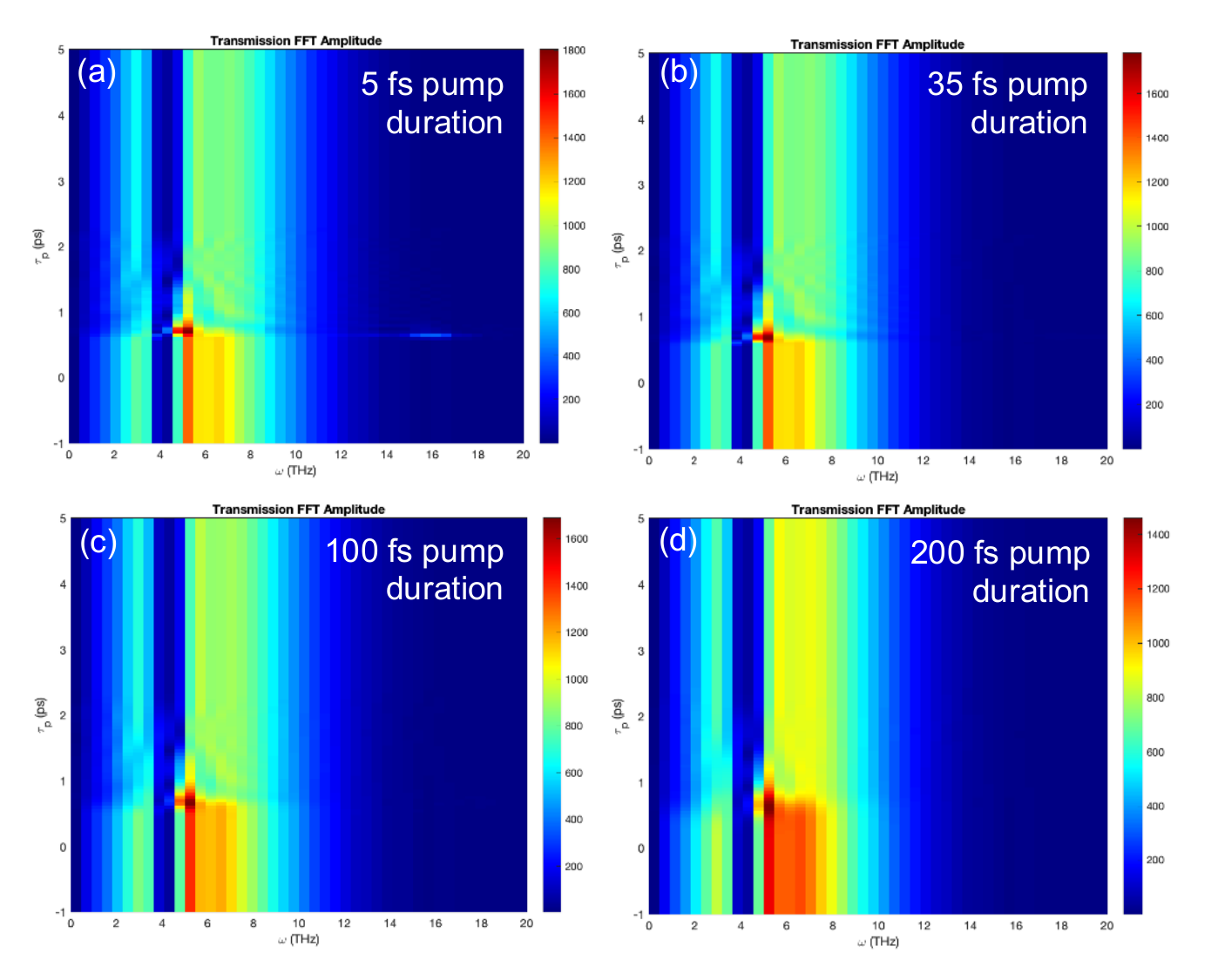}
    \caption{(a-d) FDTD transmission results showing the dependence of time-frequency features on the pump pulse duration. The systematic features evolving after the pump arrival at $\tau_p$ = 0.5~ps are suppressed for pump pulses longer than 100~fs, and are not significantly enhanced for pump pulses shorter than 35~fs.}
    \label{Asfig1}
\end{figure}

\newpage

\indent The effect of pump intensity was also considered, with select results shown in Supplemental Figure 2. The enhancement following $t = \tau_p$ scales with input pump fluence, expected if the intensity of the material parameter modulation is proportional to the input fluence, as is the case in SnSe. Here lower pump fluences are used to demonstrate the feature scaling, as it is difficult to see with stronger oscillations in these simulations. This scaling matches the experimental THz time-time data shown in Supplemental Figure 3, where the enhancement is not distinguishable at lower incident fluences (0.1 mJ/cm$^2$) and clear at higher incident fluences (7.5 mJ/cm$^2$).

\begin{figure}[ht!]
    \centering
    \includegraphics[width=0.9\textwidth, trim={7cm 3cm 8cm 3cm},clip]{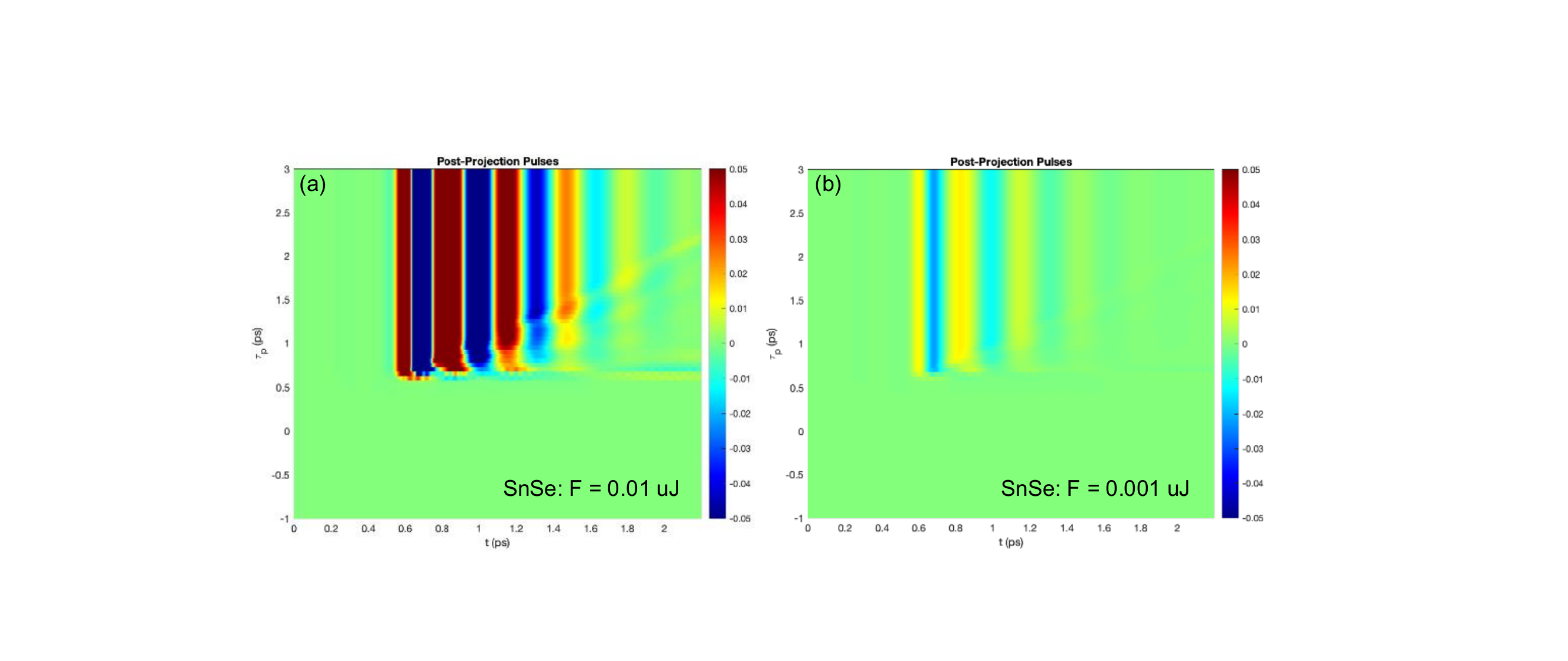}
    \caption{Projected time-time maps of $\Delta E$ at low (a) and high (b) pump fluences. While the $t = \tau_p$ response extension is present even for lower fluences, the intensity of the feature is seen to scale with excitation fluence. This matches the experimental results on SnSe shown in Supplemental Figure 3.}
    \label{Asfig2}
\end{figure}

\begin{figure}[ht!]
    \centering
    \includegraphics[width=\textwidth, trim={6.5cm 0.5cm 7.5cm 2cm},clip]{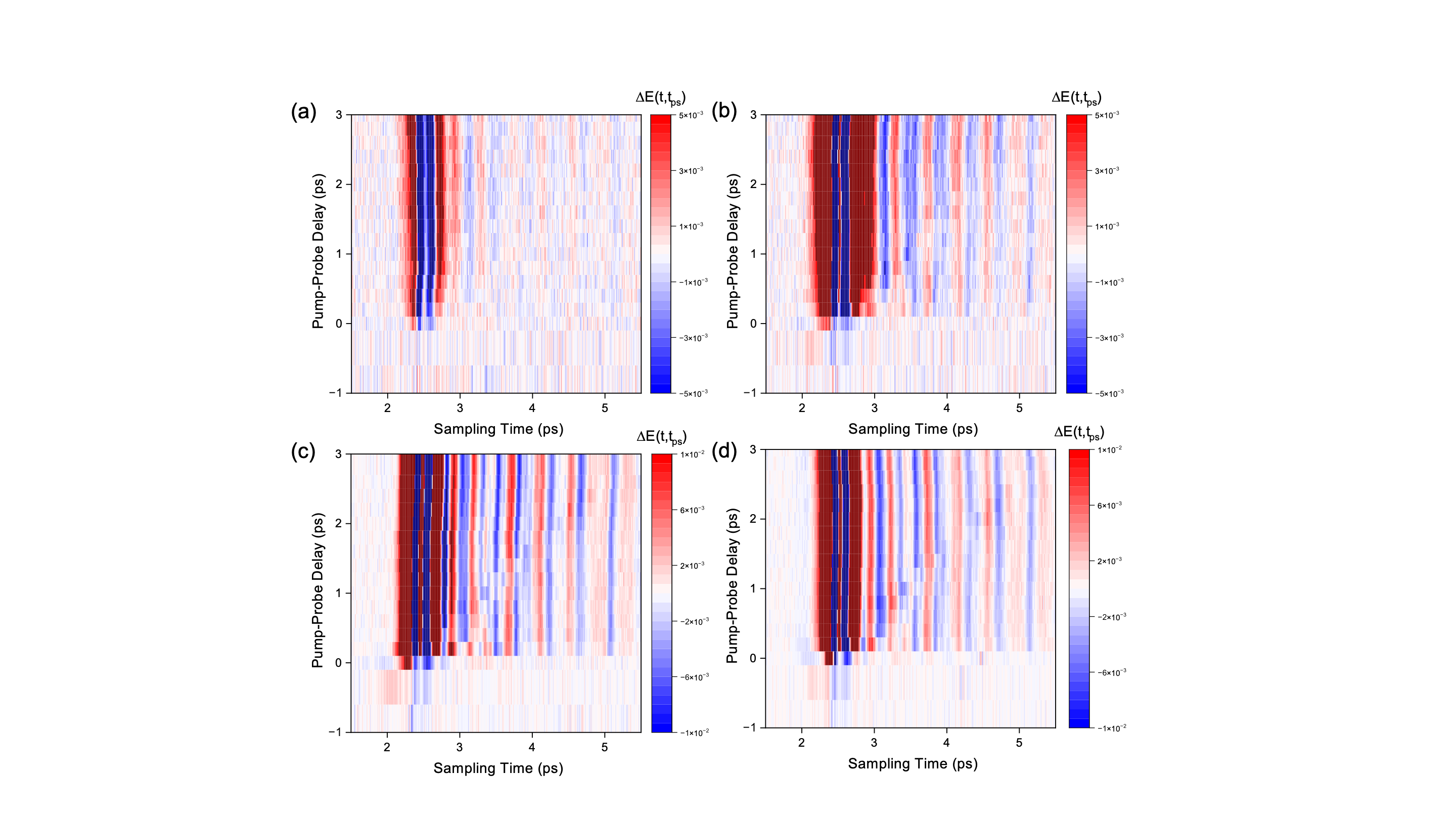}
    \caption{Experimental time-time $\Delta E$ maps for (a) 0.1 mJ/cm$^2$, (b) 1.3 mJ/cm$^2$, (c) 3.1 mJ/cm$^2$, (d) 0.1 mJ/cm$^2$. The response extension following the $t = t_{ps}$ line is seen to scale with the excitation fluence.}
    \label{Asfig3}
\end{figure}

\clearpage

\noindent The parameter dependence of the derived non-equlibrium Lorentzian conductivity was also tested, with results shown in Supplemental Figure 4. As in the FDTD simulations, the center of the response follows the Lorentzian center frequency $\omega_0$. Additionally, when increasing either the photoexcited charge lifetime or the Lorentzian linewidth (representative of the momentum relaxation time), the time frequency features are strongly suppressed. This confirms that the Lorentzian nonequilibrium formalism follows similar scaling relations to the Drude case, and should be relevant for other long-lived, narrow linewidth excitations.

\begin{figure}[ht!]
    \centering
    \includegraphics[width=0.7\textwidth, trim={5cm 0.5cm 12cm 0cm},clip,angle=-90]{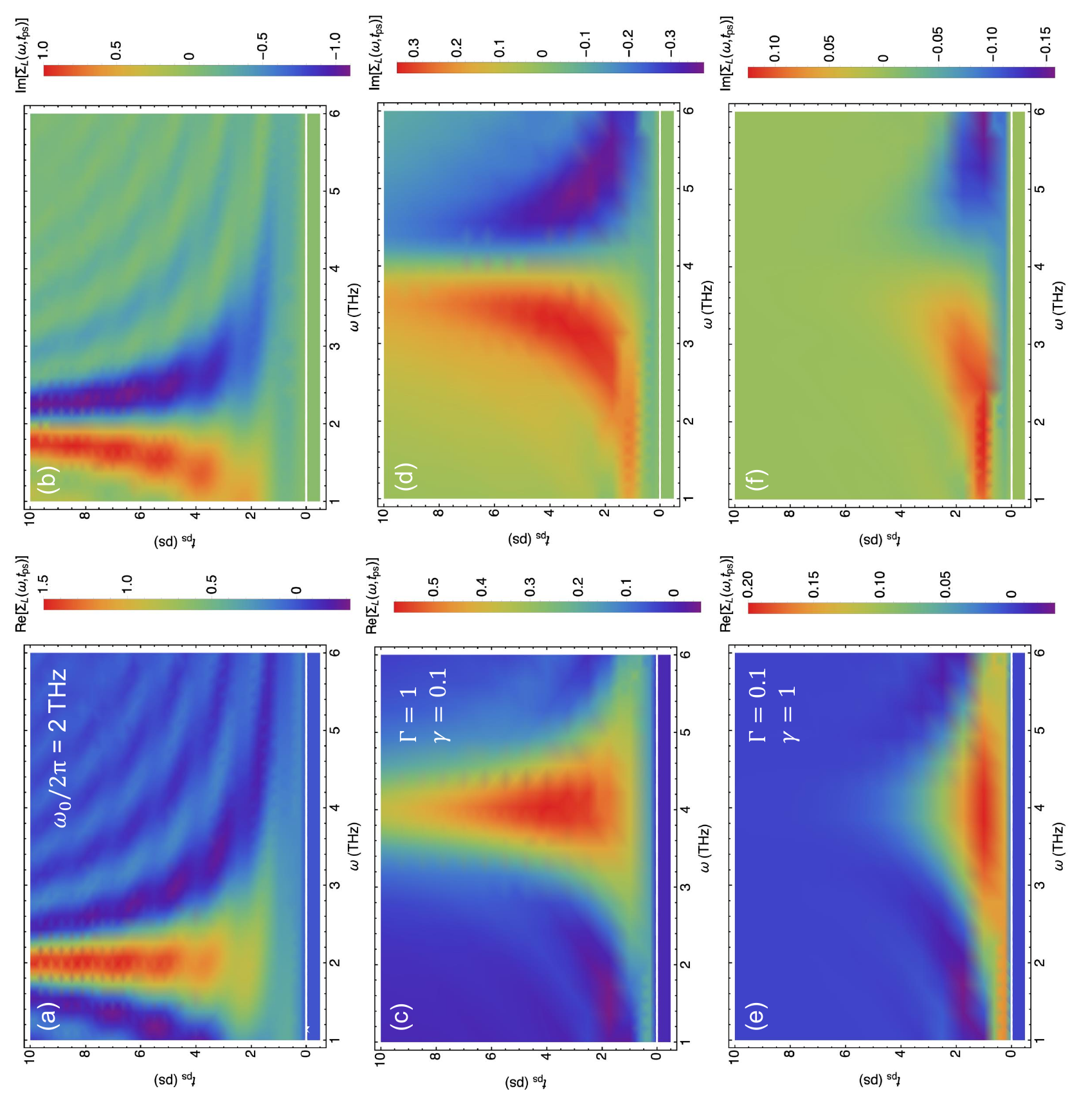}
    \caption{Various maps of the $\Sigma_L$ real and imaginary response with varied material parameters. (a,d) Response with an adjusted center frequency $\omega_0$. (b,e) Response with increased $\Gamma$. (c,f) Response with increased $\gamma$. These results confirm that the expectations from the base Drude nonequilibrium model hold for $\Sigma_L$.}
    \label{Asfig4}
\end{figure}

\clearpage

Finally, the dependence of the systematic distortion with sample temperature was investigated, as temperature can strongly affect the momentum scattering time of carriers in the material. Testing SnSe with similar pump fluences and time delays at room temperature, we note a strong suppression of time-frequency features at times soon after photoexcitation. This is ascribed to the broadening of the dispersion with temperature lowering the momentum relaxation time. Supplemental Figure 5 shows the transmission spectrum for pump fluences of 3.1 and 7.5 mJ/cm$^2$ at room temperature.

\begin{figure}[ht!]
    \centering
    \includegraphics[width=0.45\textwidth, trim={1cm 1cm 5cm 1cm},clip]{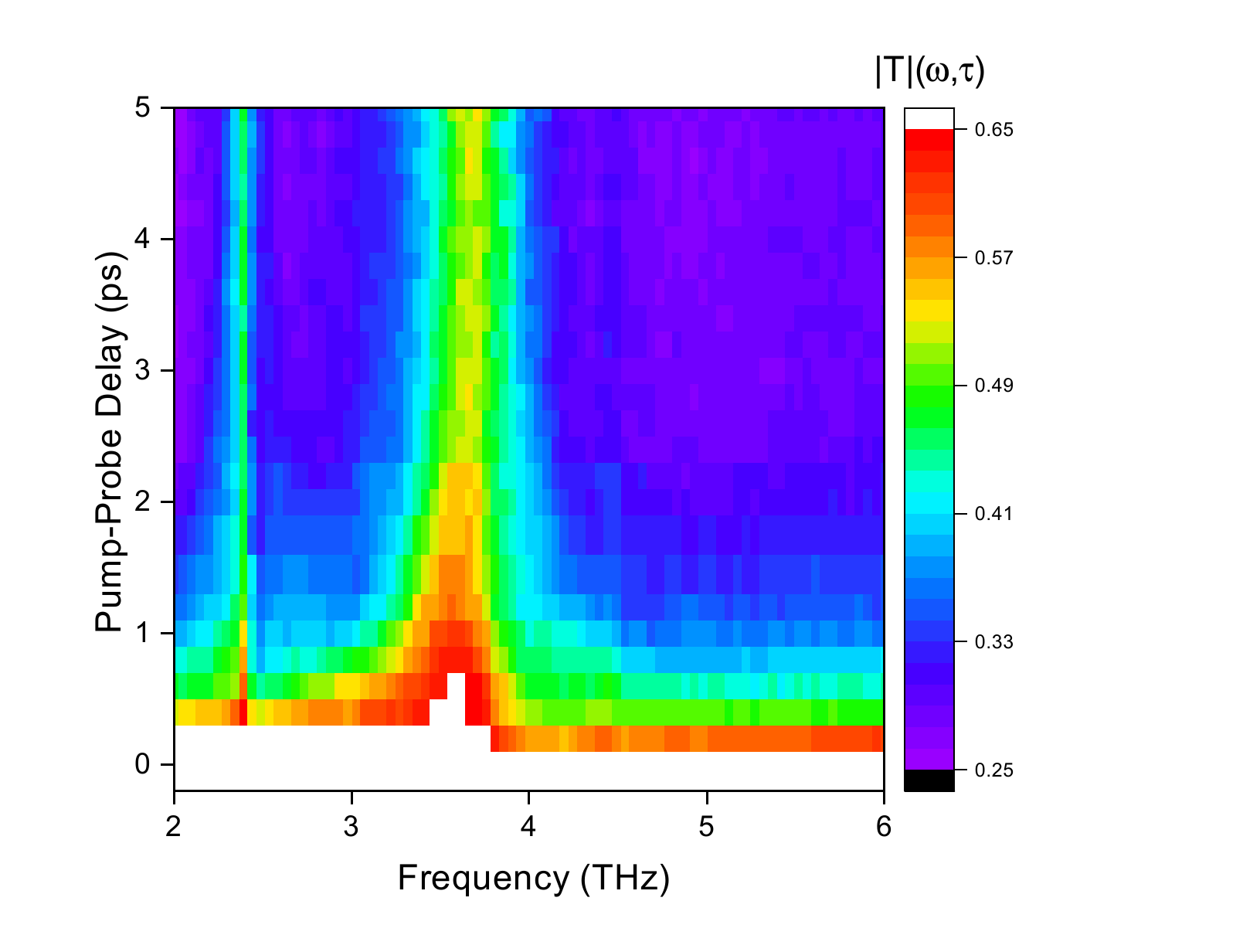}
    \includegraphics[width=0.45\textwidth, trim={1cm 1cm 5cm 1cm},clip]{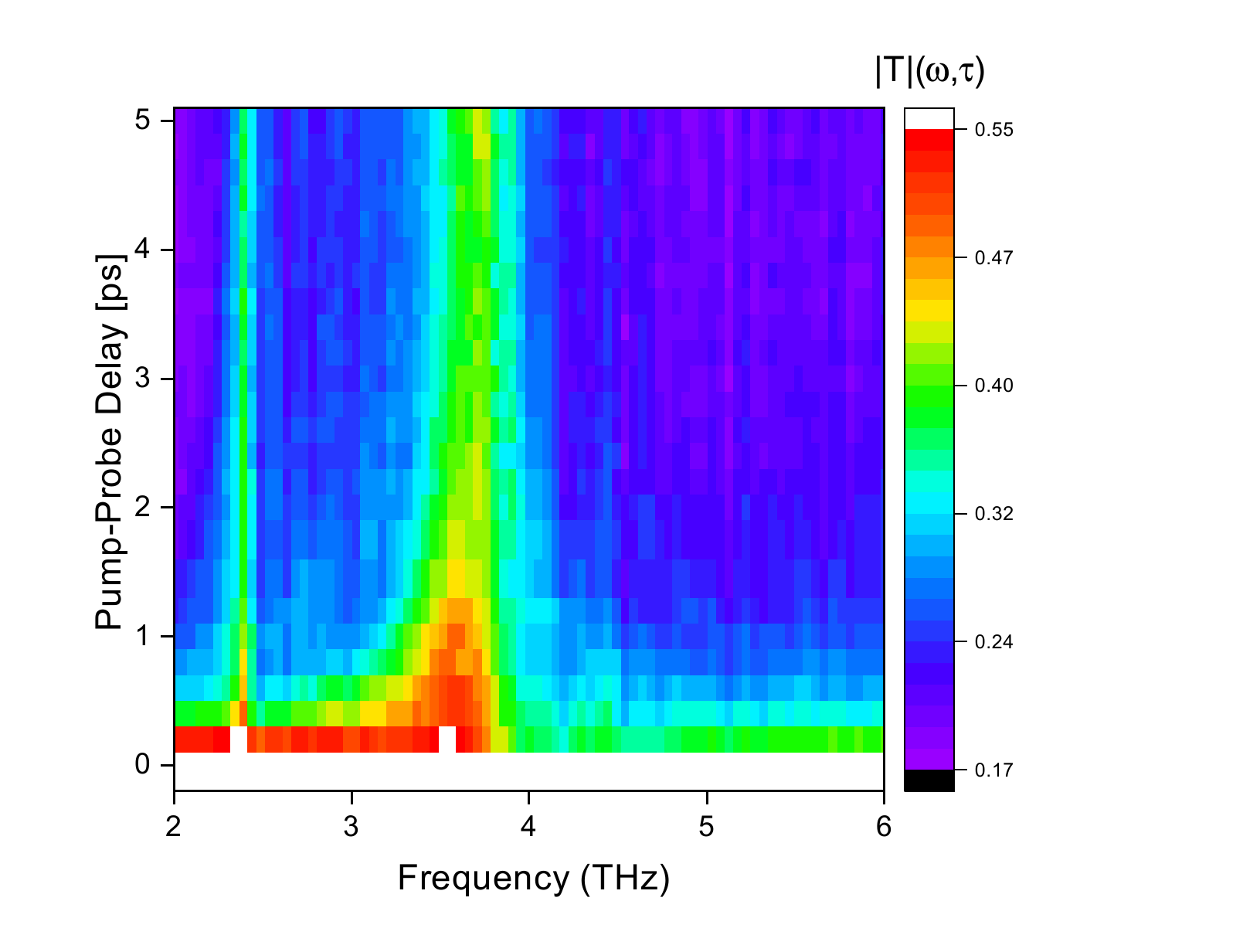}
    \caption{Transmission spectra 2D maps for 3.1 (left) and 7.5 (right) mJ/cm$^2$ pump fluences. The main phonon peak is notably broader than the tests at 80~K, and the clear time-frequencies from the low temperature 2D maps are absent.}
    \label{Asfig5}
\end{figure}
